\documentclass[aps, pra,superscriptaddress, twocolumn,showkey]{revtex4-1}
\usepackage{comment}
\usepackage{enumerate}
\usepackage{amssymb}
\usepackage{amsmath}
\usepackage{graphicx}
\usepackage{tikz}

\usepackage[colorlinks,bookmarks=false,citecolor=blue,linkcolor=green,urlcolor=blue]{hyperref}

\def\doi{http://dx.doi.org/}

\begin{document}
\title{Diffusion and operator entanglement spreading}
\author{Vincenzo Alba}
\address{Institute  for  Theoretical  Physics, Universiteit van Amsterdam,
Science Park 904, Postbus 94485, 1098 XH Amsterdam,  The  Netherlands}

\begin{abstract}
Understanding the spreading of  
the {\it operator space entanglement entropy} ($OSEE$) is key in order to 
explore out-of-equilibrium quantum many-body systems. 
Here we argue that for integrable models the 
dynamics of the $OSEE$ is related to the diffusion of the operator front. 
We derive the logarithmic bound $1/2\ln(t)$ for the $OSEE$ of some 
simple, i.e., low-rank, diagonal local operators. 
We numerically check that the bound is saturated in the rule $54$ chain, 
which is representative of interacting integrable systems.  
Remarkably, the same bound is saturated in the 
spin-1/2 Heisenberg $XXZ$ chain. Away from the 
isotropic point and from the free-fermion point, the $OSEE$ 
grows as $1/2\ln(t)$, irrespective of the chain anisotropy, 
suggesting universality. 
Finally, we discuss the effect of integrability breaking. 
We show that strong finite-time effects are present, 
which prevent from probing the asymptotic behavior of the $OSEE$. 

\end{abstract}

\maketitle

\section{Introduction}

Understanding operator spreading in quantum 
many-body systems poses several intriguing challenges. 
Given an initially local-in-space operator ${\mathcal O}$, 
its dynamics under a many-body Hamiltonian $H$ 
is ${\mathcal O}(t)=e^{i H t}{\mathcal O}e^{-iH t}$. 
The support of the operator increases with time, and 
the initially local information spreads within an emerging lightcone. 
The most urging question is as to whether  a generic local operator admits an 
efficient representation as a Matrix Product 
Operator~\cite{zwolak-2004,verstraete-2004,hastings-2006,prosen-2007,znidaric-2008,molnar-2015} (MPO). 
An affirmative answer would suggest that it is possible to 
simulate operator spreading with classical computers, with tremendous 
implications for Noisy Intermediate-Scale Quantum~\cite{preskill-2018} 
(NISQ) computing technologies. 
A figure of merit for the MPO-simulability is the so-called {\it Operator 
Space Entanglement Entropy} (OSEE), which is the entanglement entropy in 
operator space. 

Since its inception~\cite{zanardi}, 
the $OSEE$ is attracting flourishing 
interest~\cite{zanardi,znidaric-2008,molnar-2015,dubail-2017,pizorn-2009,hartmann-2009}. 
It has been suggested in Ref.~\onlinecite{prosen-2007} that 
in integrable systems the $OSEE$ grows at most logarithmically with time, 
as it was found for free fermions~\cite{pizorn-2009}. 
Very recently, a logarithmic bound has been 
derived for the so-called rule $54$ chain~\cite{adm-2019}, which 
is believed to be representative of generic integrable systems. 
This has been checked in spin chains~\cite{adm-2019}. 
Oppositely, it has been argued that the $OSEE$ grows linearly~\cite{pizorn-2009} 
in generic systems. 
Interestingly, this linear growth is predicted by the random unitary 
scenario, which posits that universal out-of-equilibrium features 
of the $OSEE$ can be captured by replacing the evolution 
operator $e^{iHt}$  with random unitary 
gates~\cite{nahum-2017,nahum-2018,keyser-2018,jonay-2018,khemani-2018}. 
Despite all these efforts, however, the general mechanism behind the dynamics  
of the $OSEE$ is yet to be unveiled, even for integrable systems. 
This is in contrast with the entanglement of a state, 
for which  a powerful quasiparticle 
picture~\cite{calabrese-2005,fagotti-2008,alba-2016,alba-2018} 
explains the entanglement dynamics in terms of the ballistic 
motion of entangled quasiparticles. 

One goal of this paper is to show that for generic integrable systems the 
$OSEE$ reflects the diffusion of the operator front. 
Here, building on Ref.~\onlinecite{adm-2019} we provide a 
tight logarithmic bound for the $OSEE$ of some simple 
operators in the rule $54$ chain. Remarkably, the same  bound is saturated in 
the spin-$1/2$  $XXZ$ chain, at least away from the free-fermion point and 
the isotropic $XXX$ point.  
This suggests a universal relation between diffusive  and 
$OSEE$ dynamics. Finally, we numerically investigate how this scenario 
is affected by integrability-breaking interactions. 

To define the $OSEE$ $S({\mathcal O})$ we  bipartite 
the system as $A\cup B$, and consider the Schmidt decomposition of ${\mathcal O}$ as 
${\mathcal O}/\sqrt{\mathrm{Tr}
({\mathcal O}^\dagger{\mathcal O})}=\sum_i\sqrt{\lambda_i}{\mathcal O}_{A,i}\otimes{\mathcal O}_{B,i}$, 
with ${\mathcal O}_{A/B,i}$ two orthonormal bases for the operators with 
support in $A$ and $B$, and $\lambda_i> 0$ the so-called Schmidt coefficients. 
The operator entanglement is $S({\mathcal O})=-\sum_i\lambda_i\ln\lambda_i$. 
%
\begin{figure}[t]
\includegraphics[width=0.35\textwidth]{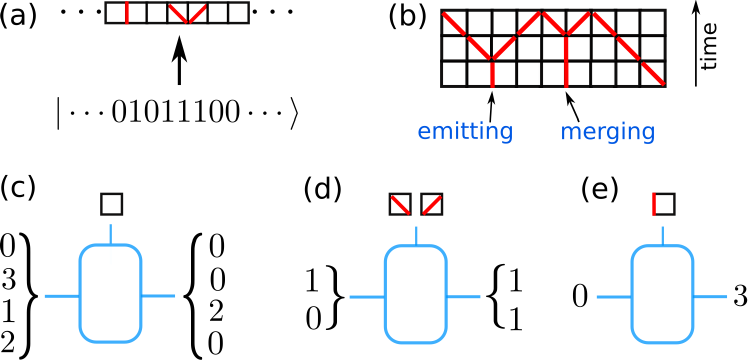}
\caption{ Rule $54$ chain. (a) Mapping to soliton space. 
 Slanted lines and vertical lines are moving solitons and scattering ones. 
 (b) Scattering solitons are time-delayed and can 
 be ``merging'' or ``emitting''. 
 (c-e) MPO representation of the identity. Large boxes denote the tensors 
 $A^{\tau_x}_{\beta_x,\beta_{x+1}}$ at site $x$. The physical index 
 $\tau_x$ (vertical leg) takes values in soliton space. 
 The allowed values of the virtual indices (horizontal legs) 
 $\beta_x,\beta_{x+1}$ are reported. 
}
\label{fig0:identity}
\end{figure}
%

%
\begin{figure*}[t]
\includegraphics[width=1.\textwidth]{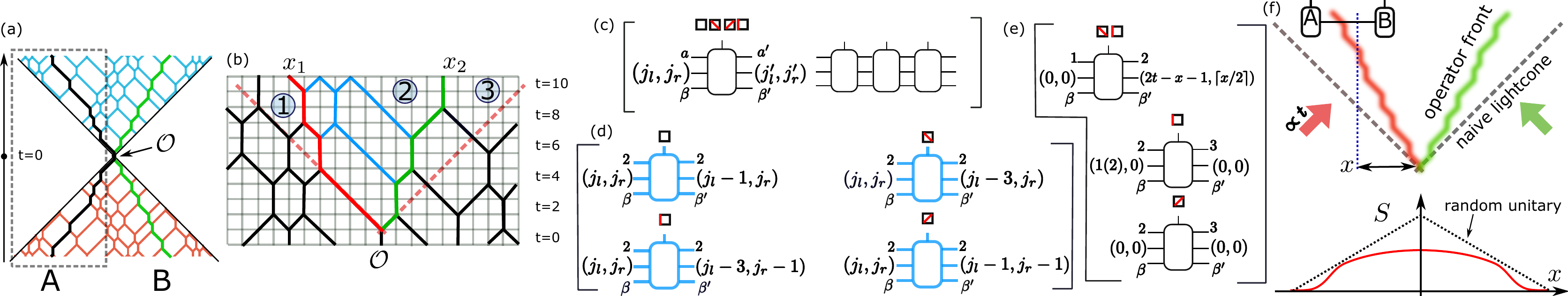}
\caption{ (a) Dynamics of a diagonal operator ${\mathcal O}$ in the rule 
 $54$. (a) Double lightcone. ${\mathcal O}(0)$ creates 
 a pair of scattering left/right movers $x=0$.  
 They scatter with the background solitons. The upper and the 
 lower half-lightcones coincide. 
 (b) Typical evolution. Solitons positions $x_1,x_2$ are measured from the 
 left edge of the ligtht-cone (dashed line). 
 Three different regions appear. Region $2$ is 
 the ``reduced lightcone''. 
 (c) MPO representation for ${\mathcal O}(t)$. Index $a=1,2,3$ 
 denotes the three regions in (b). The composite index $(j_l,j_r)$ with $j_l,j_r\in[1,t]$ 
 tracks the positions of the two solitons, and 
 $\beta,\beta'$ are as in  Fig.~\ref{fig0:identity}. 
 In $1$ and $3$, ${\mathcal O}$ is the identity, and 
 $a=a'=1,3$ and $j_l=j_l'=j_r=j_r'=0$. 
 An example of MPO contraction is shown. 
 (d) MPO in region $2$. All nonzero tensor elements are shown. (e) 
 Tensors at the interface between different regions. 
 At $1,2$ one has  a left mover and $(j_l,j_r)=(2t-x-1,\lceil x/2\rceil)$.  
 At $1,3$ the right mover that emerged at the center is found. 
 (f) Cartoon for OSEE spreading in integrable systems (top). 
 The operator front spreads with the dressed velocity $v_d$, implying that 
 a number $\propto t$ of left and right moving solitons are present in the lightcone. 
 The bipartition as $A\cup B$ with $A=[-t,x]$ is shown. The OSEE reflects 
 the number of ways of distributing between $A$ and $B$ the solitons that are 
 present in the lightcone. The effective MPO describing ${\mathcal O}$ is reported. 
 The virtual indices of the grouped tensors for $A$ and $B$ take values 
 $\mathrm{min}(t-x,t+x)$, corresponding to the maximum number of 
 solitons that can be stored in the smaller of the two subsytems. 
 As a result the entanglement profile as a function of 
 the cut position exhibits a ``pancake'' structure (see bottom), as opposed with the 
 random-unitary scenario, which gives a ``pyramid'' profile (dotted profile). 
}
\label{fig1:bobenko}
\end{figure*}
%
\section{$OSEE$ in the rule $54$ chain}  
\label{sec:rule54}

Here we focus on the $OSEE$ spreading in the rule $54$ chain~\cite{bobenko-1993}. 
The Hilbert space is that of a system of qubits $s_x=0,1$. 
The dynamics is generated by a three-site unitary 
gate $U_{x}$ acting as   

\begin{equation}
\label{eq:gate}
U_x=|s_{x-1},s'_x,s_{x+1}\rangle\langle s_{x-1}s_x,s_{x+1}|, 
\end{equation}
where $s'_x=s_{x-1}+s_{x+1}-s_{x-1}s_{x+1}$. $U_x$ flips the 
qubit at $x$ if one of the neighbouring qubits is $1$. 
Any qubit configuration is evolved as 
$U=\prod_{\mathrm{even}\, x}\prod_{\mathrm{odd}\,x}U_x$. 
The rule $54$ chain possesses well-defined quasiparticles, which is the key property of 
integrable systems. Quasiparticles are 
emergent left/right moving solitons. They correspond to pairs of 
adjacent qubits that are in the $1$ state (more details are reported in 
Appendix~\ref{sec:app-3}). Crucially, solitons undergo 
pairwise elastic scattering, which is implemented as a Wigner 
time delay~\cite{wigner-1955} (cf.~Fig.~\ref{fig0:identity}). 
Again, this is also generic for integrable models (see 
Ref.~\onlinecite{doyon-2018}). Two 
solitons that are scattering correspond to the qubit configuration $010$. 
The mapping between qubits and left/right movers is encoded as an MPO with 
bond dimension $\chi=4$. Here 
we work directly in soliton space. 
As it is shown in Figure~\ref{fig0:identity}, a 
site $x$ can be empty (empty box), or occupied by a left (right) 
mover (boxes with slanted lines in the Figure) if $(-1)^{x+t}=-1(1)$,  or by 
two scattering solitons (vertical lines). If  $(-1)^{x+t}=-1$ the two 
(``emitting'') solitons will reappear at time $t+1$, whereas if $(-1)^{x+t}=1$ 
the (``merging'') solitons will reappear at $t+2$, reflecting the 
Wigner delay. 
We are interested in the Heisenberg dynamics of local operators. 
Let us first consider the identity operator  $\mathbf{1}=
\prod_{x=0}^{L-1}\sum_{s_x}|s_x\rangle\langle s_x|$
in soliton space. 
As for all diagonal operators, one can consider the evolution of the ket or bra 
separately, because they evolve in the same way under application of $U$ and $U^\dagger$. 
One now has the evolution of the ``flat'' superposition 
$\prod_x\sum_{s_x}|s_x\rangle$. In soliton space this maps to the flat superposition 
of all allowed soliton configurations. This is efficiently encoded as an MPO (see Appendix~\ref{sec:app-3})
as $\mathbf{1}=\sum_{\{\beta_x\}}
\prod_i A^{\tau_x}_{\beta_x,\beta_{x+1}}|\tau_x\rangle\langle\tau_x|$. 
Here $A^{\tau_x}_{\beta_x,\beta_{x+1}}$ is a tensor living on site $x$. 
The index $\tau_x$ labels the soliton configuration,  $\beta_x\in [0,\chi]$ 
are the virtual indices, with $\chi$ the bond dimension. Here $A^{\tau_x}_{\beta_x,\beta_{x+1}}=1$ 
only for the cases shown in Fig.~\ref{fig0:identity} (c-e),  and 
it is zero otherwise. 
The role of $\beta_x$ is to enforce some kinematic constraints, 
for instance, that a left mover is followed only by a right mover or by 
an empty site (see Fig.~\ref{fig8:sol} in Appendix~\ref{sec:app-3}). 
Since $\chi$ is small and the identity operator does not evolve, one has 
that $S(\mathbf{1})$ is constant in time. 

This changes dramatically for the $OSEE$ of a local operator. 
By adapting a remarkable result of Ref.~\onlinecite{klobas-2019} it has 
has been shown~\cite{adm-2019} that 
the dynamics of operators 
is described  by an MPO with $\chi\propto t^2$. This implies 
the ``naive'' bound $S({\mathcal O})\le 2\ln(t)$ for the $OSEE$. Here 
we show that  the growth of the $OSEE$  reflects the fluctuations of the 
number of solitons between $A$ and its complement. 
This 
allows us to derive a tighter bound for the $OSEE$ spreading. 
To derive our result, we review the construction of the MPO for 
the {\it diagonal} operator that inserts two scattering 
solitons at $L/2$, i.e., ${\mathcal O}=|010\rangle\langle 010|$. 
This is illustrated in Fig.~\ref{fig1:bobenko}. ${\mathcal O}$ is diagonal, 
implying that the upper and the lower lightcones coincide. At $t>0$ a left 
and right movers are emitted from $L/2$. They play a crucial role in 
the MPO contruction. Indeed, ${\mathcal O}(t)$ 
corresponds to the flat superposition of all the possible soliton configurations that 
contain the left and right movers that were inserted at the origin at $t=0$. 
This simple constraint on the soliton configurations 
implies that the $OSEE$ grows logarithmically. 

We note that as the solitons emitted from the center scatter with the 
background solitons, they undergo two biased random walks. 
Their positions $x_1,x_2=0,1,\dots$ at time $t$, which are 
measured from the left edge of the lightcone (dashed lines in 
Fig.~\ref{fig1:bobenko} (b)), are determined by the scatterings. 
The crucial observation is that 
all the background solitons that scattered with the 
two solitons emitted from the center are contained in 
the ``reduced lightcone'' within them (region $2$ in 
Fig.~\ref{fig1:bobenko} (b)). Outside of the 
reduced lightcone ${\mathcal O}(t)$ is the identity. 
To construct the MPO for ${\mathcal O}(t)$ we complement the 
MPO for the identity in Fig.~\ref{fig0:identity} with some extra indices. First, we 
introduce an index $a=1,2,3$ to keep track of the different regions. 
The number of left/right movers in region $2$ is tracked 
by two extra indices $j_l,j_r$. Finally, the index $\beta$ 
is as in Fig.~\ref{fig0:identity}. 
The structure of the MPO is summarised in Fig.~\ref{fig1:bobenko} (c). 
Physically, $j_r$ at $(x,t)$ counts the number of right movers in region $2$, 
whereas $j_l$ is the expected distance between $x$ and the 
right mover that emerged from the center, assuming that there are 
no left movers in the remaining interval $[x+1,t]$ of the lightcone. 
In regions $1,3$ we set  $a=a'=1,3$, and $j_l=j_l'=j_r=j_r'=0$. 
The allowed values of $j_l,j_r$ in region $2$ for which the MPO is nonzero 
are reported in Fig.~\ref{fig1:bobenko} (d).  The  interpretation is straightforward. 
For instance, if at site $x$ there is no soliton, one has 
$j'_r=j_r$ and $j'_l=j_l-1$, because 
the distance from the right mover emerged from the center decreases by one after 
moving to the next site. If at $x$ there is a right mover, 
then $j_r'=j_r-1$. If a left mover is present, one has that $j_l'=j_l-3$  because 
the left mover shifts the right mover emerging from the center by two 
sites to the left (see Fig.~\ref{fig1:bobenko} (b)). 
Finally, Fig.~\ref{fig1:bobenko} (e) shows the tensors at the 
interface between regions $1,2$ and $2,3$. At the boundary $1,2$ a left mover 
is present, and $j_l,j_r$ is initialized as 
$j_r=\lceil x_1/2\rceil$ and $j_l=2t-x_1-1$. 
At the boundary $2,3$ one has $j_l=j_l'=0$, ensuring  that all 
the background solitons expected within the reduced lightcone 
have been found and the right mover that emerged from the center is on that site. 
Notice that there is a subtlety due to the kinematics 
of solitons if two scattering solitons are met at $2,3$ (see Fig.~\ref{fig1:bobenko} (e)). This, however, 
does not affect the leading logarithmic growth of the $OSEE$. 
Now, since $0\le j_l,j_r\le t$, the bond dimension of the MPO that describes ${\mathcal O}(t)$ 
is clearly $\chi\propto t^2$, implying that $S\le2\ln(t)$. 
To proceed, we observe that due to the scatterings, the left and right 
movers that emerged from 
the center move with a ``dressed'' velocity~\cite{gopala-2018,sarang-2018} 
$v_d=1/2$ (the bare velocity is $v_b=1$). Crucially, their 
trajectories, and the operator front, exhibit diffusion~\cite{jacopo,sarang-2018}. 
This diffusion is essential to have nonzero entanglement. 
Indeed, the dressed solitons behave as free particles, their   
trajectories cross each other. This implies that a flat superposition of 
dressed solitons is mapped onto itself by the dynamics, which implies 
the absence of entanglement production. 

We now observe that in the reduced lightcone there are $\propto t$ left/right movers.  
Let us consider the bipartition $A\cup B=[-t,x]\cup[x,t]$, with $x\le0$. 
A crude approximation for ${\mathcal O}(t)$ gives 
\begin{equation}
\label{eq:decomp}
{\mathcal O}(t)=\sum_{k=0}^{t-|x|}\frac{\sqrt{\binom{t-|x|}{k}
\binom{t+|x|}{t-k}}}{\sqrt{\binom{2t}{t}}}
{\mathcal O}^A_k\otimes{\mathcal O}^B_{t-k}. 
\end{equation}
Here ${\mathcal O}^{A}_k$ and ${\mathcal O}^B_{t-k}$ are normalised operators 
in $A$ and $B$ constructed with $k$ and $t-k$ solitons. In~\eqref{eq:decomp} we 
assume that ${\mathcal O}_{k}^{A}$ and ${\mathcal O}_{t-k}^{B}$ are some 
``flat'' superpositions of all the configurations with $k$ and $t-k$ solitons, i.e., 
we assume that the positions of the background solitons are maximally ``scrambled'' 
within the reduced lightcone. 
This is not true in general because solitons 
scatter locally. We also assume that ${\mathcal O}_k^{A}$ and ${\mathcal O}_{t-k}^{B}$ 
form orthonormal bases for $A$ and $B$. 
The two binomials in the sum in~\eqref{eq:decomp} give the number of ways of 
arranging the  solitons in the two subsystems. Note that for large $t$ the behavior of~\eqref{eq:decomp}  
is dominated by the configurations with $k=(t-|x|)/2$, showing a 
spreading $\sqrt{t}$. This reflects that there is an average number $(t-|x|)/2$ of solitons 
in subsystem $A$. The number of solitons in $A$ fluctuates, the fluctuations being 
$\propto\sqrt{t}$.  We anticipate that these fluctuations are responsible 
for the growth of the $OSEE$. 
Crucially, this mechanism is different from the 
spreading of the state entanglement after a global quantum quench, where 
entanglement is produced locally at each point in space and it is transported 
by entangled multiplets of 
quasiparticles~\cite{calabrese-2005,fagotti-2008,alba-2016}. 
This is also different from the random-unitary scenario. The main assumption 
of this scenario is that the entanglement profile $S(x,t)$ satisfies the 
equation $\partial_t S=\Gamma(\partial_x S)$. Here $\Gamma$ is the entropy 
production rate, which depends on the spatial variation of the entropy profile, 
and it is nonzero at any point in space. 
This implies that the entanglement profile 
has the typical ``pyramid'' shape (Fig.~\ref{fig1:bobenko} (f)). 
In contrast, the logarithmic growth 
in integrable systems is reflected in a ``pancake'' structure in the entanglement profile (Fig.~\ref{fig1:bobenko} (f), 
see also Appendix~\ref{sec:app-2}). 


We can now derive a bound on the $OSEE$ growth from~\eqref{eq:decomp}. 
The bond dimension of the decomposition 
from~\eqref{eq:decomp} is $t-|x|+1$. Note that $t-|x|$ is the largest number of 
solitons that can be accomodated within $A$. The eigenvalues of the reduced density 
matrix for $A$ are simply 
$\lambda_k=\binom{t-|x|}{k}\binom{t+|x|}{t-k}/\binom{2t}{t}$. Notice that 
the fact that there are only $\propto t$ eigenvalues is an approximation. 
In the rule $54$ chain one should expect $\propto t^2$ nonzero eigenvalues, 
instead of the $\propto t$ predicted by the argument above. 
On the other hand, the number $\propto t^2$  does not imply the scaling $2\ln(t)$ for the $OSEE$ 
because the eigenvalues are not equal but exhibit a nontrivial distribution. 
By using the explicit form of $\lambda_k$ one obtains the analytical 
bound for the  $OSEE$  as (see Ref.~\onlinecite{kiefer-2020} for a similar calculation) 
\begin{equation}
\label{eq:bound}
S_\mathrm{max}= \frac{1}{2}\ln(t). 
\end{equation}
%
%
\begin{figure}[t]
\includegraphics[width=0.475\textwidth]{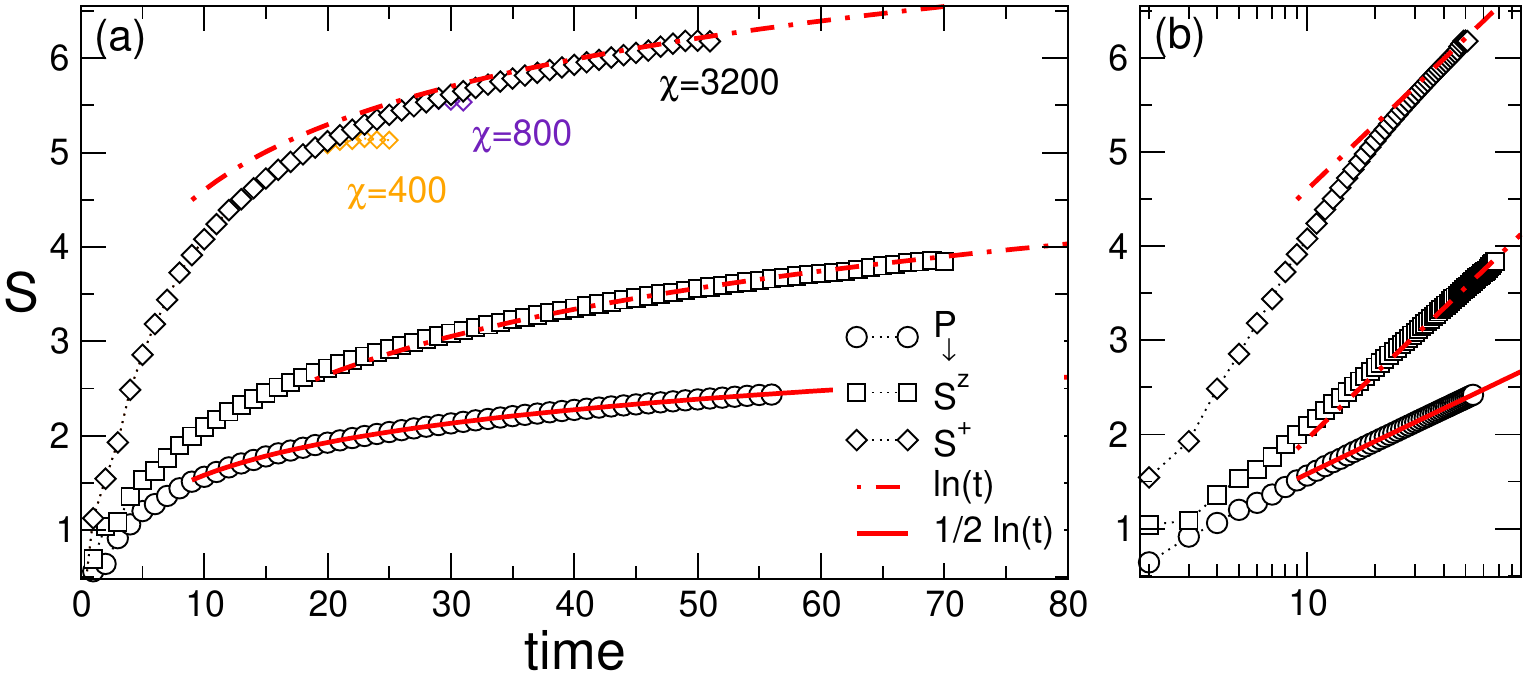}
\caption{ (a) Operator entanglement in the rule $54$ chain. Different symbols are 
 $tDMRG$ data for $P_\downarrow\equiv(\mathbf{1}/2-S^z)$ (circles), 
 $S^z$ (squares), and $S^+$ (diamonds), inserted at the chain center. 
 The continuous and dashed-dotted 
 lines are $S_\mathrm{max}=1/2\ln(t)$, and $2S_\mathrm{max}$, respectively. 
 (b) Same data as in (b) using a logarithmic scale on the $x$-axis. 
}
\label{fig2:bobenko}
\end{figure}
%
Crucially, the prefactor $1/2$ in~\eqref{eq:bound} is reminiscent of the 
$\sqrt{t}$ fluctuations in the number of solitons in the subsystems $A$ and $B$. 
Eq.~\eqref{eq:bound} is expected to hold for the simple, i.e., low-rank, 
diagonal operator. We should remark that the prefactor of the 
$OSEE$  growth should depend on the structure of the operator. For instance, 
the identity operator, for which the 
$OSEE$ is constant in time, is $\mathbf{1}=
P_\downarrow+P_\uparrow$. On the other hand, the $OSEE$ of 
$S_z=P_\uparrow-P_\downarrow$ grows logarithmically. Moreover, 
(see Ref.~\onlinecite{jonay-2018}) for traceful 
operators, the prefactor of the $OSEE$ growth depends on the 
trace. 
Also, for off-diagonal operators (see Fig.~\ref{fig1:bobenko} (a)) 
the upper and lower lightcones do not coincide, suggesting a faster growth 
of the $OSEE$. 

We should also stress that the behavior of the $OSEE$ 
in free-fermion systems is different from~\eqref{eq:bound}. For instance, 
the $OSEE$ of $S^z$ saturates, whereas 
that of $S^x$ increases as $1/3\ln(t)$. Interestingly, 
the prefactor $1/3$ could reflect the absence of 
diffusion for free fermions, suggesting that the $OSEE$ could be 
potentially useful to distinguish interacting integrable from free 
systems~\cite{spohn-2018}. 
%
\section{Integrable dynamics: rule $54$ and $XXZ$ spin chain} 

To benchmark our main result~\eqref{eq:bound}, in Fig.~\ref{fig2:bobenko} we discuss 
the case of the rule $54$ chain. 
We focus on the projector operator 
$P_\downarrow\equiv(\mathbf{1}/2-S^z)$, the raising operator 
$S^+$, and $S^z$, all inserted at the center of the chain. 
The symbols are $tDMRG$ data~\cite{uli,paeckel-2019,itensor}. 
For $S^+$, we report the bond dimension $\chi$. 
The full line is Eq.~\eqref{eq:bound}, whereas the dashed-dotted line 
is $2S_\mathrm{max}$. 
The agreement between~\eqref{eq:bound} and the data is excellent for $P_\downarrow$, 
signalling that the bound~\eqref{eq:bound} is saturated. 
For $S^z$ one should also expect $S=2S_\mathrm{max}$ 
(see Ref.~\onlinecite{jonay-2018}). 
A fit to $\kappa\ln(t)+a$ gives $\kappa\approx0.9$. 
For $S^+$, we observe a reasonable agreement with $2S_\mathrm{max}$, 
although finite-time effects seem larger. 

We now discuss the universality of~\eqref{eq:bound}. 
We consider a generalisation of the spin-$1/2$ $XXZ$ chain 
defined by the Hamiltonian 
\begin{multline}
H=\sum_{i=1}^{L}\frac{1}{2}(S^+_iS^-_{i+1}+S_i^-S_{i+1}^+)
\\+\Delta\sum_{i=1}^{L}S_i^zS_{i+1}^z+\Delta'\sum_{i=1}^LS_i^z S_{i+2}^z
\end{multline}
where $\Delta,\Delta'$ are  real parameters. For $\Delta'=0$ the model is integrable for 
any $\Delta$, whereas $\Delta'\ne0$  breaks integrability (see Appendix~\ref{sec:app-1}). 
%
\begin{figure}[t]
\includegraphics[width=0.475\textwidth]{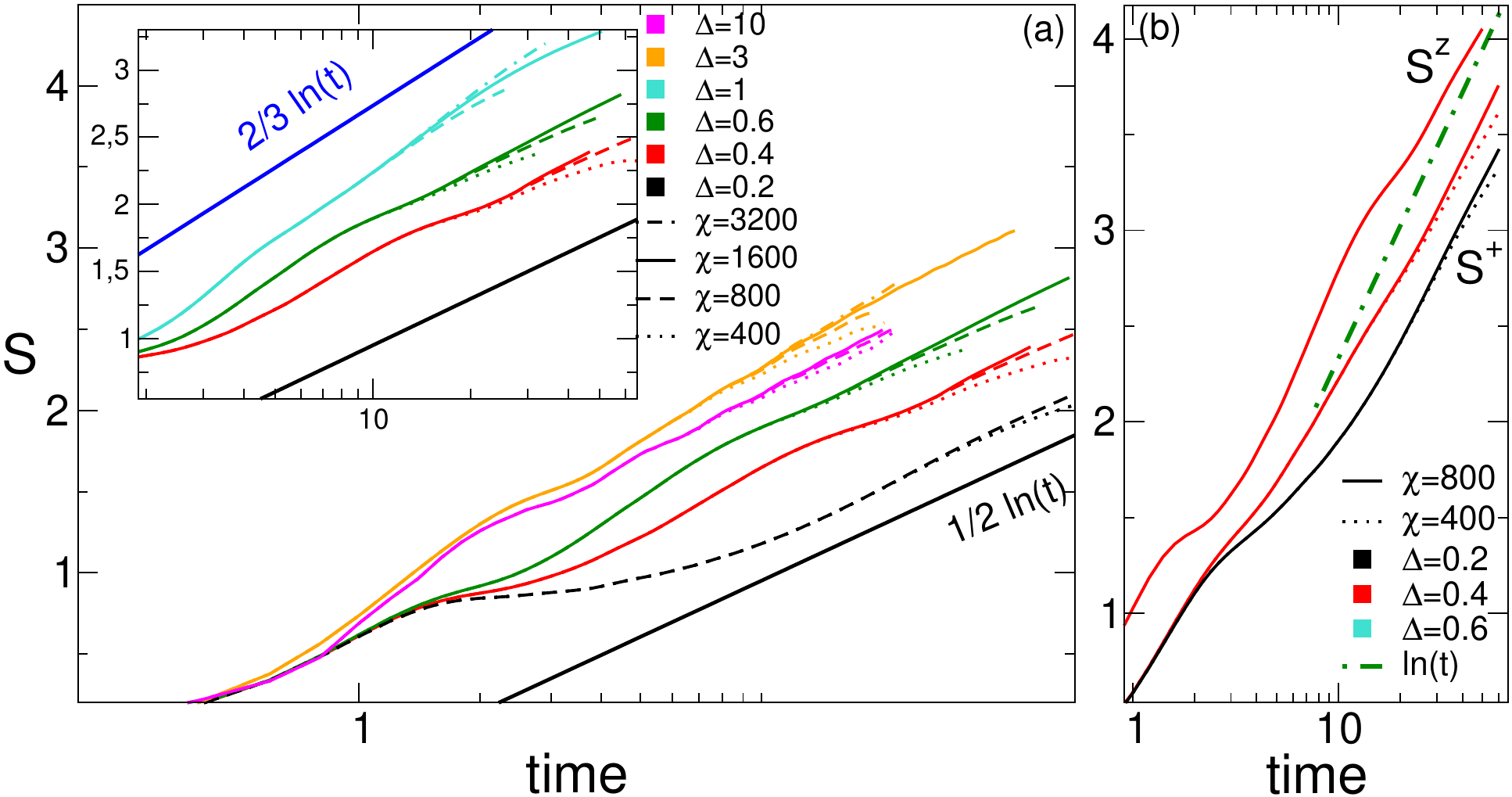}
\caption{$OSEE$ in the $XXZ$ chain. (a) tDMRG data 
 for $P_\downarrow\equiv (\mathbf{1}/2-S^z)$ for several values of 
 $\Delta$. Different line styles are different bond dimensions 
 $\chi$. For $\Delta\ne1$ data are compatible with $S_\mathrm{max}$. 
 For $\Delta=1$ (inset) the $OSEE$ shows a faster growth. 
 (b) Dynamics of the $OSEE$ of $S^+$ and $S^z$. 
}
\label{fig3:xxz}
\end{figure}
%
Let us consider the integrable case, i.e., $\Delta'=0$. 
We discuss the $OSEE$ of $P_\downarrow$ 
in Fig.~\ref{fig3:xxz} (a) and that of $S^+$ and $S^z$ in Fig.~\ref{fig3:xxz} (b). 
The $tDMRG$ data for $P_\downarrow$ exhibit a clear logarithmic increase. For 
$\Delta\ne1$ they are compatible with $S_\mathrm{max}+c(\Delta)$,  
suggesting universality of the prefactor $1/2$ of the $OSEE$. 
Interestingly, $c(\Delta)$ reflects the behavior of the 
diffusion constant~\cite{ilievski-2018,gopala-2019}, i.e., it increases with $\Delta$ 
for $0\le \Delta\le1$, then 
it decreases for $\Delta>1$, saturating for $\Delta\to\infty$. 
For $\Delta\to1$ the diffusion constant diverges~\cite{ilievski-2018,gopala-2019}, 
which signals superdiffusive behavior, suggesting violations of~\eqref{eq:bound} for $\Delta=1$. 
The data in the inset of Fig.~\ref{fig3:xxz} might suggest 
the behavior  $\kappa\ln(t)$ with $k>1/2$, 
although they could just signal large finite-time corrections.  
It has been proposed in Ref.~\onlinecite{ljubotina-2017} 
that the superdiffusive behavior as $t^{2/3}$ arises at $\Delta=1$, 
suggesting $S\propto 2/3\ln(t)$ (reported for comparison in 
Fig.~\ref{fig3:xxz}). 
Finally, in Fig.~\ref{fig3:xxz} (b) we discuss $S^z$ and 
$S^+$. The $OSEE$ increases faster. Finite-time 
effects are large, and the evidence for the 
behavior $S\propto 2S_\mathrm{max}$ is weak. 

%
\begin{figure}[t]
\includegraphics[width=0.45\textwidth]{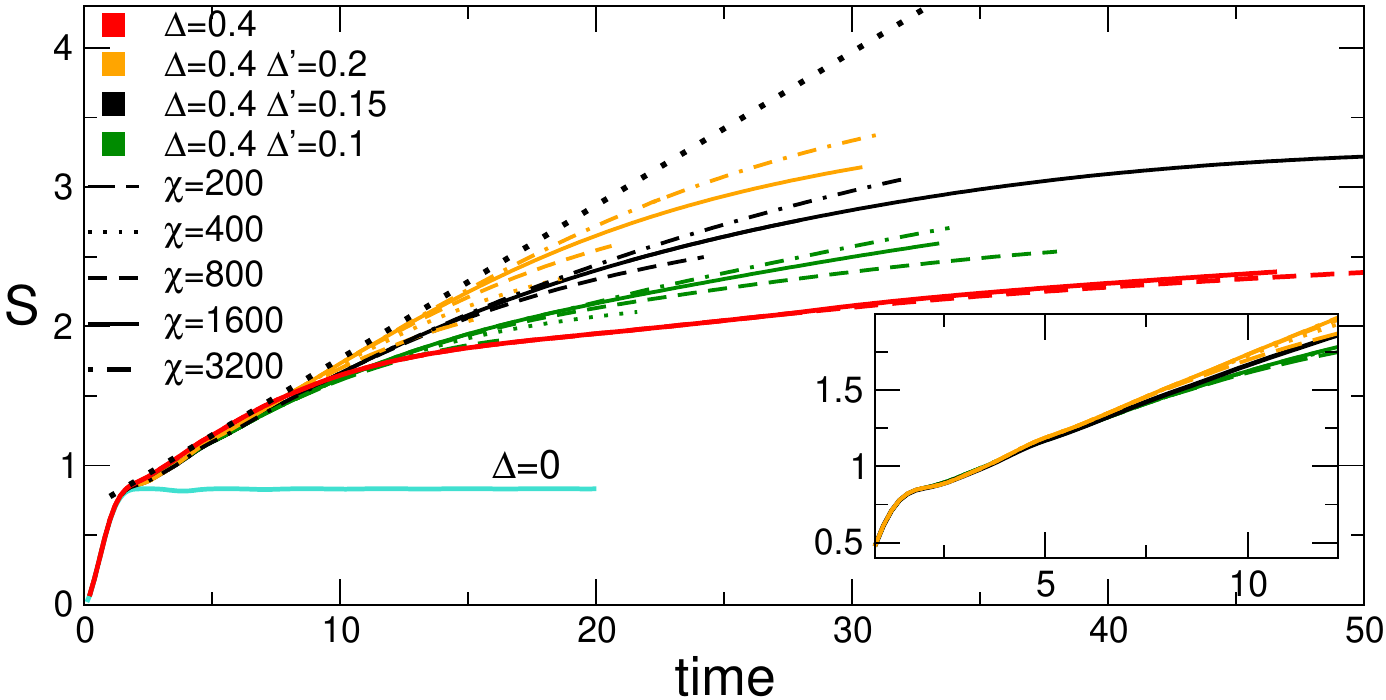}
\caption{ $OSEE$ dynamics and  integrability breaking: 
 $OSEE$ of $P_\downarrow\equiv(\mathbf{1}/2-S^z)$ 
 inserted at the center of the chain. 
 Lines are $tDMRG$ data for the $XXZ$ chain 
 with $\Delta=0.4$ and several values of $\Delta'$. Different bond 
 dimensions $\chi$ are shown. We also show the data for 
 $\Delta=\Delta'=0$, i.e., the free-fermion point, where the $OSEE$ 
 saturates. The dotted line is $S\simeq 0.11 t+0.66$, and it is obtained by fitting the 
 approximately linear behavior at intermediate times. The inset is a zoom for $t\le 10$. 
}
\label{fig3:chaos}
\end{figure}
%

\section{Non-integrable dynamics} 
\label{sec:nonint}

The soliton picture should 
breakdown for generic models, because they do not possess quasiparticles. 
According to the random-unitary scenario, this would imply a linear growth of the $OSEE$. 
However, it has been suggested in Ref.~\onlinecite{muth-2011} that if a 
 conservation law is present, the R\'enyi operator entanglement $S^{(2)}$ 
 of the associated local operator exhibits logarithmic growth, even if the 
 system is nonintegrable. 
Notice that for systems without conservation laws, for 
instance Floquet systems, the linear growth of operator entanglement 
is supported by exact calculations~\cite{bertini-2019,bertini-2020,bertini-2020a}. 
Our $tDMRG$ results for $P_\downarrow$ are in Fig.~\ref{fig3:chaos}. It is enlightening to  
first consider the integrable case for $\Delta'=0$ and $\Delta=0.4$. 
At very short times $t\approx 2$, the $OSEE$  exhibits a jump, reflecting that at 
$\Delta=0$ the $OSEE$ saturates (see the result for $\Delta=0$ in the Figure). 
Then, there is an intermediate regime, where 
a nearly-linear growth is present. The asymptotic 
behavior sets in at longer times. 
Upon breaking integrability $tDMRG$ simulations 
become more challenging. 
At short times a linear increase is observed. However, this could be  
reminiscent of the transient regime also observed for $\Delta'=0$.  
In fact, a change in behavior 
happens at $t^*(\Delta')$, with $t^*$ increasing with $\Delta'$. 
The data in Fig.~\ref{fig3:chaos} are compatible 
with two scenarios. In one scenario the $OSEE$ increases linearly at 
asymptotically long times. The asymptotic  regime sets in after 
a long transient  in which the system behaves as if it was integrable. 
The prefactor of the linear growth should presumably increase with $\Delta'$. 
Alternatively, the breaking of integrability  gives rise to a longer transient, 
as compared with the integrable case, before the logarithmic behavior 
sets in. Longer transients should be expected 
generically for nonintegrable systems because transport is dominated by 
diffusion. 

\section{Conclusions} 
We have shown that in integrable systems the growth of the $OSEE$ of some simple 
operators exhibits a logarithmic increase. 
Our work opens several research avenues. First, it would important to derive {\it ab initio} 
the behavior in~\eqref{eq:bound}, at least in the rule $54$ chain, for instance, by using the 
recent developments in Ref.~\onlinecite{klobas-2020,klobas-2020a}. It is also important 
to understand the $OSEE$ for more complicated operators and systems. 
Our data for non-integrable systems do not allow to reach a conclusion on the 
behavior of the $OSEE$ in generic systems, although they are compatible with Ref.~\onlinecite{muth-2011}. 
It is of fundamental importance 
to clarify this issue. 
Finally, the argument leading to~\eqref{eq:bound} 
gives that $S_\mathrm{max}$ is the same for all the R\'enyi entropies $S^{(\alpha)}$. 
However, we numerically checked that although $S^{(\alpha)}$ exhibit logarithmic growth, 
the prefactor is smaller than $1/2$ and it depends on $\alpha$. 
It would be interesting to clarify this issue by studying the R\'enyi 
entropies. Finally, it would be interesting 
to clarify the relationship between $OSEE$ and anomalous transport, for instance 
superdiffusion~\cite{gopala-2019,dupont-2020,jacopo-1}.

\begin{acknowledgments}
I am grateful to Jerome Dubail and Marko Medenjak for introducing me 
to the problem of operator entanglement and to the Bobenko chain, and 
for several important discussions. I would also like to thank Maurizio 
Fagotti and Bruno Bertini for several discussions. 
I acknowledge support from 
the European Research Council under ERC Advanced grant 743032 DYNAMINT. 
\end{acknowledgments}

\appendix
\section{Spectral diagnostic for the non-integrable case}
\label{sec:app-1}

Here we address the integrability of the hamiltonian~\eqref{eq:xxz-ni}. 
We consider the general hamiltonian 
\begin{equation}
\label{eq:xxz-ni}
H=H_\mathrm{XXZ}+
\sum_{i}\frac{J'}{2}(S^+_iS^-_{i+2}+S_i^-S_{i+2}^+)
+\Delta'\sum_{i}S_i^zS_{i+2}^z, 
\end{equation}
where $H_{XXZ}$ is the standard Heisenberg $XXZ$ hamiltonian 
\begin{equation}
\label{eq:xxz}
H_{XXZ}=
\sum_{i}\frac{1}{2}(S^+_iS^-_{i+1}+S_i^-S_{i+1}^+)
+\Delta\sum_{i}S_i^zS_{i+1}^z. 
\end{equation}
For $J'=\Delta'=0$ one recovers the $XXZ$ chain, which is integrable by 
the Bethe ansatz for any $\Delta$. To understand the effect of the integrability 
breaking terms we study the gaps $\delta_n$ between adjacent levels of the 
energy spectrum of~\eqref{eq:xxz-ni}. Here we define $\delta_n$ as 
\begin{equation}
\delta_n\equiv E_{n+1}-E_n, 
\end{equation}
with $E_n$ energy  levels. 
For chaotic systems the behavior of $\delta_n$ should be described by an 
appropriate random matrix ensemble, provided that the contribution of the  
density of states, which is model dependent, is removed. An alternative solution is 
to focus on the ratio between consecutive gaps $r_n$ as~\cite{oganesyan-2007}.  
\begin{equation}
\label{equ:ratio}
0\le r_n\equiv\mathrm{min}\{\delta_n,\delta_{n-1}\}/\mathrm{max}\{\delta_n,\delta_{n-1}\}\le1. 
\end{equation}
For Poisson-distributed energy levels, i.e., for integrable systems, 
the average value of the ratio is $\langle r_n\rangle=2\ln(2)-1\approx0.386$. In the 
non-integrable case one should expect that energy levels are described 
by the Gaussian Orthogonal Ensemble ($GOE$). This gives[47] 
$\langle r_n\rangle=4-2\sqrt{3}\approx0.535$.  
%
\begin{figure}[t]
\includegraphics[width=0.4\textwidth]{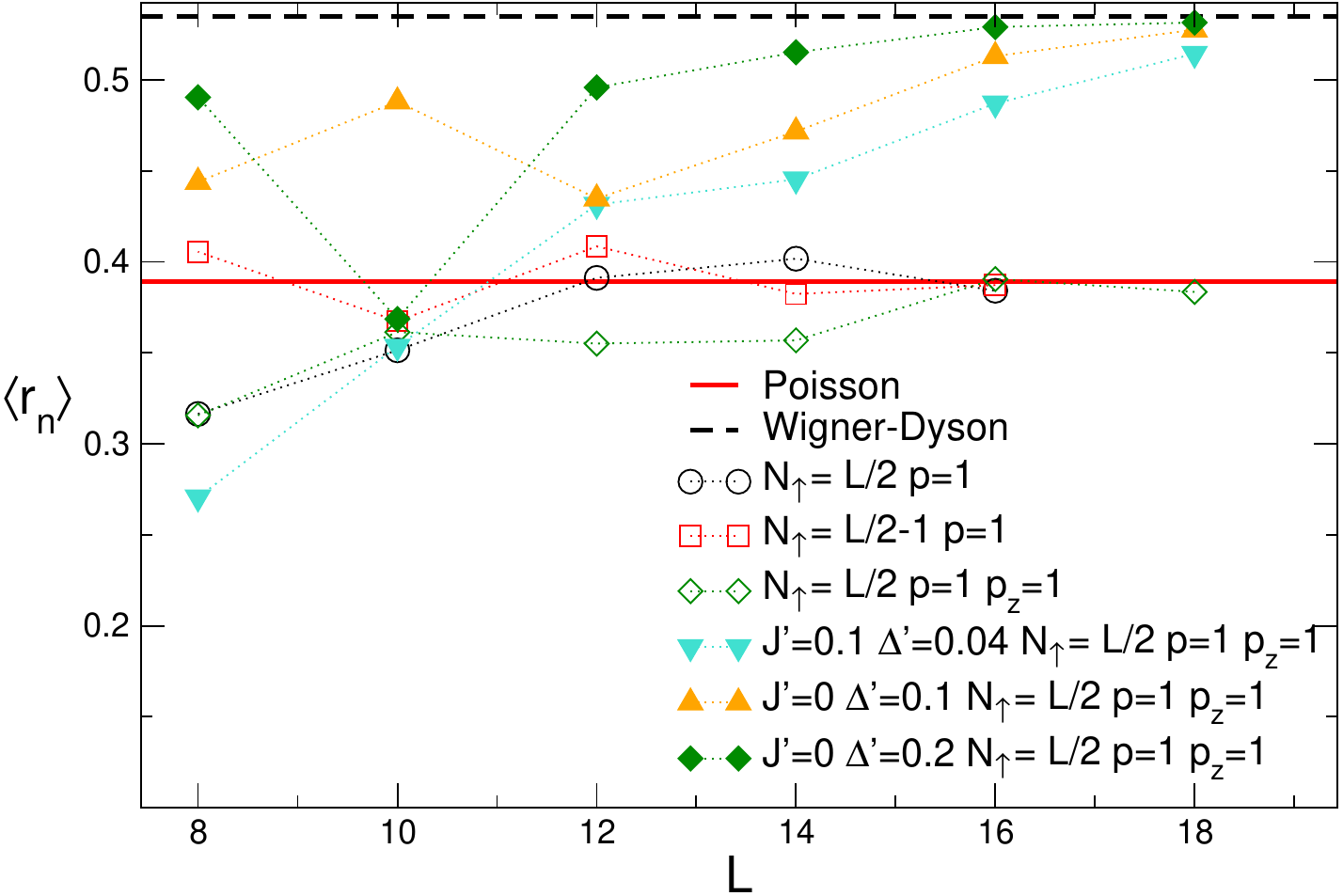}
\caption{ Spectral diagnostics of integrability-breaking. The figure shows the 
 ratio of consecutive gaps $r_n$ (cf.~\eqref{equ:ratio}) versus the system size 
 $L$ for the hamiltonian~\eqref{eq:xxz-ni}. Here we focus on the 
 case with $\Delta=0.4$, and several $J'$ and $\Delta'$. The empty symbols are 
 the data for the integrable case $J'=\Delta'=0$. 
 The different symbols correspond to different number of up spins (magnetization) 
 $N_\uparrow$, spatial parity eigenvalue $p$, and spin inversion eigenvalue 
 $p_z$. Only the quantum numbers that are fixed are reported in the legend. 
 The full and dashed lines are the expected results for integrable and 
 chaotic models. 
}
\label{fig4:gaps}
\end{figure}
%
Our results are reported in Fig.~\ref{fig4:gaps}. The data are obtained from 
exact diagonalisation of a chain with $L\le 18$ sites. Periodic boundary 
conditions are used. In the Figure $N_\uparrow$ is the number of up spins, which 
fixes the magnetization sector. Most of the data are at half-filling $N_\uparrow=L/2$, 
although we consider also $N_\uparrow=L/2-1$. We denote with $p=\pm1$ the eigenvalue 
of the parity under reflection with respect to the center of the chain. Here $p_z=\pm1$ is 
the eigenvalue of the spin inversion operator. Empty symbols are for the integrable 
case, i.e., the $XXZ$ chain with $\Delta=0.4$ (cf.~\eqref{eq:xxz}). The 
different symbols are for different symmetry sectors. In the legend we only 
report the quantum numbers that are fixed. The results for the integrable case 
are reasonably close to the expected value $\langle r_n\rangle\approx 0.386$, 
at least in the limit $L\to\infty$. 

This is different upon breaking integrability. The data are reported as full symbols 
in Fig.~\ref{fig4:gaps}. First, one should stress that 
the Wigner-Dyson result $\langle r_n\rangle\approx 0.535$ is expected to hold in 
the limit $L\to\infty$ if one factors out all the conserved quantities. 
The down-triangle in the figure are the data for $J'=0.1$ and $\Delta'=0.04$. 
Clearly, finite-size corrections are present, although the data for the largest 
size $L=18$ are converging to the expected result. The up triangles and the 
diamonds are the data for $J'=0$ and $\Delta'=0.1$ and $\Delta'=0.2$, respectively. 
Upon increasing $\Delta'$, the data 
approach the Wigner-Dyson result faster, as expected. Still, in both cases there 
is reasonable agreement with the random matrix result for $L=18$.  
However, we should remark that, although the analysis performed here suggests that 
for $\Delta'=0.1,0.2$ the hamiltonian~\eqref{eq:xxz-ni} is not integrable, 
it does now give any information on the time-scale after which the effect of 
the integrability-breaking interactions start to appear. 

\section{Entanglement profiles}
\label{sec:app-2}
%
\begin{figure}[t]
\includegraphics[width=0.4\textwidth]{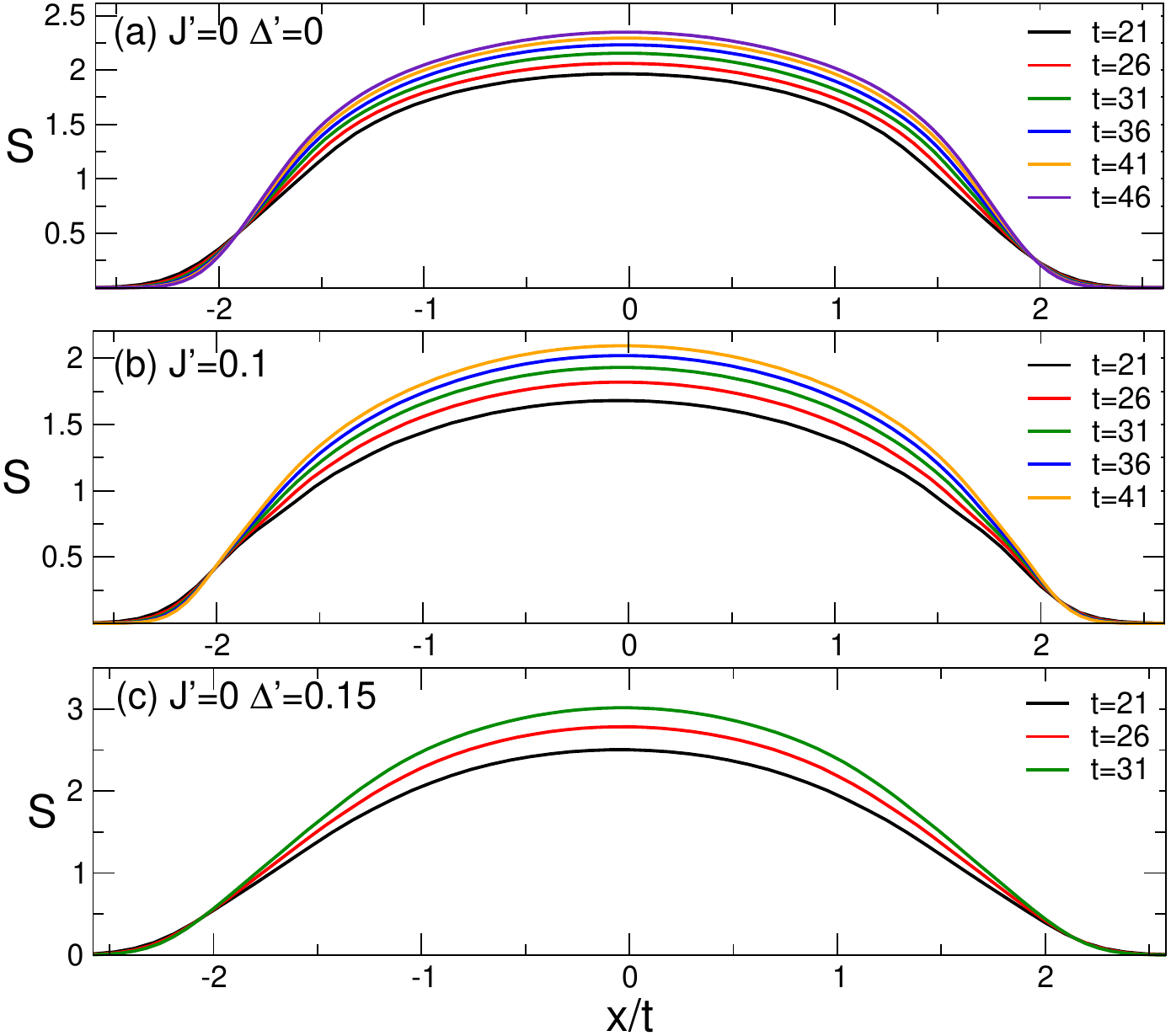}
\caption{ Profile of the operator entanglement. The results are for $P_\downarrow
 \equiv \mathbf{1}/2-S^z$ inserted at the center of the chain. The operator entanglement 
 is plotted as a function of the rescaled position $x/t$, with $x$ measured 
 from the chain center and $t$ the time. (a) shows the integrable case, i.e., 
 the $XXZ$ chain with $\Delta=0.4$. In (b) and (c) we consider the nonintegrable 
 deformation of the $XXZ$. 
}
\label{fig5:profile}
\end{figure}
%
Here we discuss the behavior of the spatial profile of the $OSEE$ of the projector operator 
$P_\downarrow\equiv \mathbf{1}/2-S^z$ in both integrable and non-integrable systems.  
The operator is inserted at the center of the chain. 
Our results are presented in Fig.~\ref{fig5:profile}. We consider the deformed XXZ chain 
hamiltonian in~\eqref{eq:xxz-ni}. We fix $\Delta=0.4$. In Fig.~\ref{fig5:profile} (a) 
we focus on the integrable case $J'=0$ and $\Delta'=0$. The figure shows the $OSEE$ plotted 
as a function of $x/t$, with $x$ the distance from the center of the chain. 
Clearly, outside of the lighcone for $|x/t|>2$ the $OSEE$ 
vanishes. Within the lightcone, in the integrable case the entanglement profile exhibits a rather 
flat behavior. This is in contrast with the expected behavior in generic systems described 
by random unitaries, for which the $OSEE$ has a maximum at $x=0$ and decreases linearly 
with the distance from the center, exhibiting a ``pyramid-like'' structure. 

In Fig.~\ref{fig5:profile} (b) we consider the effect of the integrability breaking. We now 
fix $J'=0.1$ and $\Delta'=0$. 
An important observation is that since we are interested in the long time limit and 
the $OSEE$ generically grows faster upon increasing the strength of the integrability-breaking terms 
we are limited to relatively weak integrability breaking. 
The entanglement profile is qualitative similar to the integrable 
case in Fig.~\ref{fig5:profile} (a). A similar behavior is observed in the case with $J'=0$ 
and $\Delta'=0.15$ (see Fig.~\ref{fig5:profile} (c)). 

\section{Solitonic machines}
\label{sec:app-3}

%
\begin{figure}[t]
\includegraphics[width=0.35\textwidth]{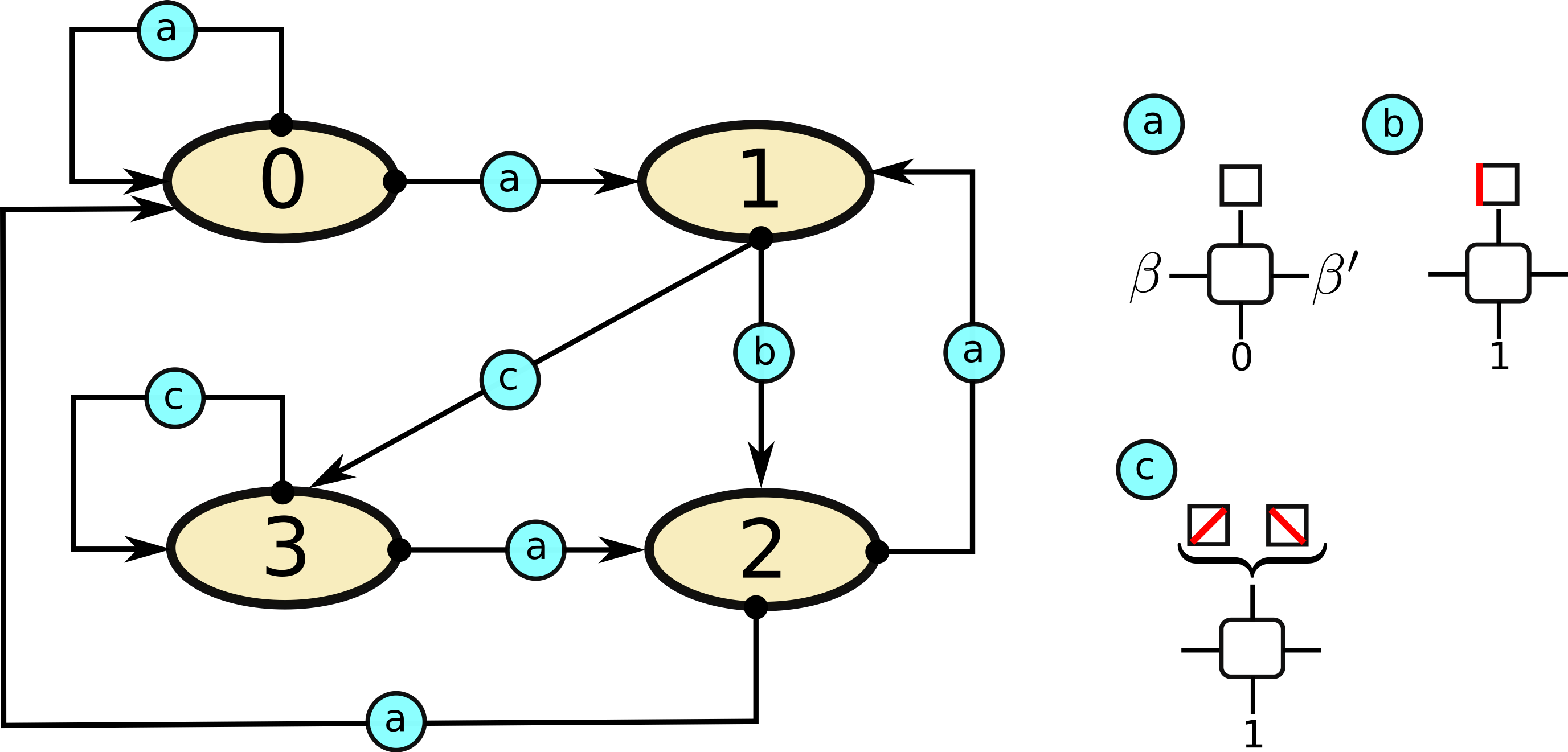}
\caption{ MPO representation of the mapping between  computational 
 basis and the soliton basis in the rule $54$ chain. The diagram shows the 
 finite-state machine encoding the mapping. The possible states of the machine 
 are labeled as $s=0,1,2,3$. The arrows denote transitions between different 
 states. (a-c) Tensors forming the MPO. The 
 lower indices take values $0,1$. The upper index 
 can be the empty box (no solitons),  slanted 
 lines denoting left and right movers, and the vertical line, which corresponds to 
 a pair of scattering solitons. The presence of the left and right mover 
 depends on the combined parity of spatial position and time. The virtual 
 indices $\beta,\beta'$ for which the tensor 
 is nonzero are the states of the machine connected by the tensor. 
}
\label{fig6:mpo}
\end{figure}
%

The mapping between the computational basis and the soliton basis for the rule $54$ chain  
(see Fig.~\ref{fig0:identity}) 
is reported in Fig.~\ref{fig6:mpo} in the framework of finite-state machines. The possible 
states of the machine are $s=0,1,2,3$. These are the states that are explored by 
a machine that scans a bit configuration site by site proceeding from left to right. 
The internal states of the machine are determined by the bit configurations on 
nearest-neighbour sites. The goal of the machine is to identify pairs of consecutive 
$11$, which correspond to left/right movers, and the configuration $010$, which 
corresponds to two scattering solitons. 
Let us assume that the machine is at site $x$ and that $s_x=s_{x-1}=0$. This defines 
the internal state $0$ of the machine. 
State $1$ means that $s_x=1$ and $s_{x-1}=0$. 
State $2$ is defined by the condition $s_x=0$ and $s_{x-1}=1$. Finally state 
$3$ means that $s_x=s_{x-1}=1$. 
All the transitions between the different states are reported in 
the diagram in Fig.~\ref{fig6:mpo}. When the machine moves from $x$ to 
$x+1$ it gives as an output the soliton configuration on $x$. For instance, 
if the transition is $0\to0$ the only possibility is that on $x$ there is no 
soliton. The possible transitions define all the nonzero elements of the 
tensors $A_{\beta,\beta'}^{s_x,\tau_x}$ forming the MPO that implements the mapping. 
Here $\beta,\beta'=0,1,2,3$ are the virtual indices of the MPO, whereas $s_x$ and  $\tau_x$ 
are the physical indices taking values in the bit space and in the soliton space, 
respectively.

%
\begin{figure}[t]
\includegraphics[width=0.35\textwidth]{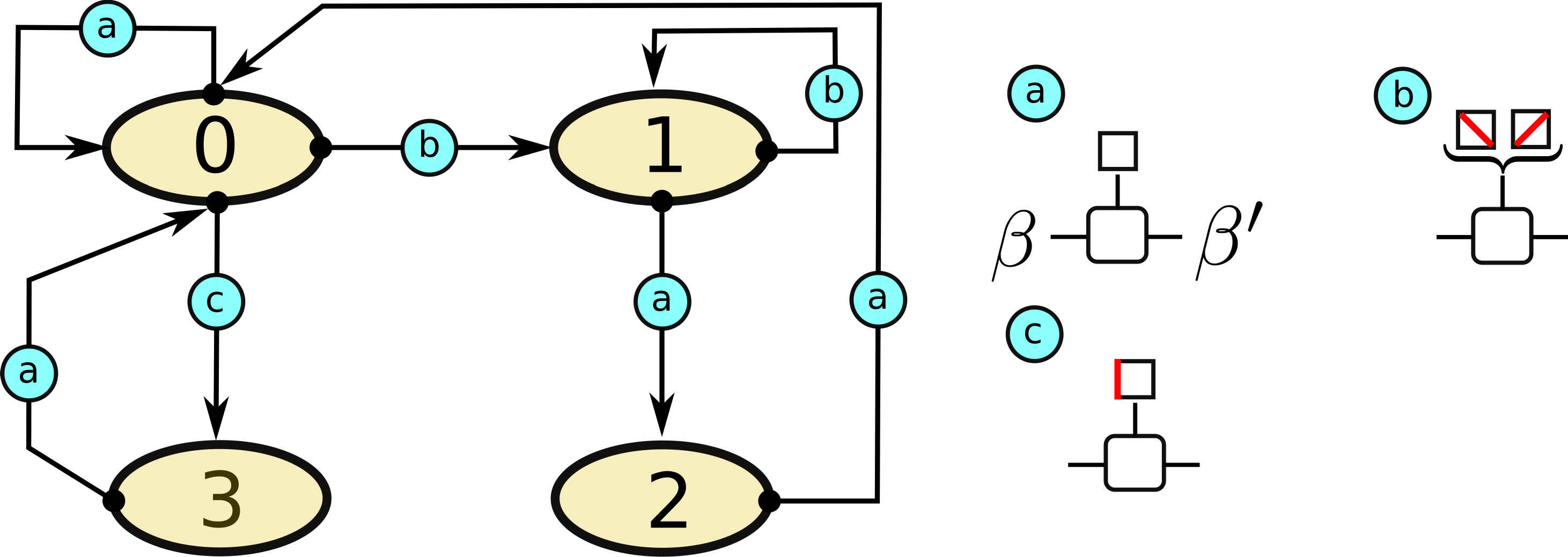}
\caption{ Soliton machine that generates the MPO representation of the identity operator 
 (infinite-temperature state). In (a-c) we report the 
 tensors forming the MPO representation. The virtual indices of the tensor have 
 values in the space of the machine states $\beta,\beta'=0,1,2,3$. 
}
\label{fig7:mpo}
\end{figure}
%

%
\begin{figure}[t]
\includegraphics[width=0.35\textwidth]{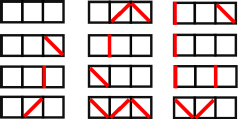}
\caption{ All the possible solitonic configurations on a system with 
 $L=3$ sites. The configurations are obtained by using the MPO representation 
 of the identity in Fig.~\ref{fig7:mpo}. 
}
\label{fig8:sol}
\end{figure}
%
From the mapping in Fig.~\ref{fig7:mpo} one can read out all the possible solitonic 
configurations on $L$ sites. The machine generating them gives the MPO 
representation of the identity operator in soliton space. 
The MPO representing the identity, or, equivalently the infinite 
temperature state in the space of solitons is shown in Fig.~\ref{fig7:mpo}. 
The meaning of the machine states is not the same as in Fig.~\ref{fig6:mpo}. 
Now state $0$ means that at site $x$ there is no solitons and on $x-1$ 
there were no free left and right movers (slanted lines). 
State $1$ means that on $x$ there is a left/right mover. 
State $2$ is defined by the condition that on $x$ there is no soliton and 
a left/right mover is present at $x-1$. 
Finally, state $3$ means that on site $x$ there is a pair of scattering 
solitons (vertical line). Note that the presence of state $2$ imposes some 
kinematic constraint for the solitons, i.e., that 
a left and right mover has to be followed by at least two empty boxes. 
To illustrate the solitonic patterns that correspond to the identity, 
in Fig.~\ref{fig8:sol} we report all the solitonic configurations that 
are allowed on three sites.

\


\begin{thebibliography}{0}%
\makeatletter
\providecommand \@ifxundefined [1]{%
 \@ifx{#1\undefined}
}%
\providecommand \@ifnum [1]{%
 \ifnum #1\expandafter \@firstoftwo
 \else \expandafter \@secondoftwo
 \fi
}%
\providecommand \@ifx [1]{%
 \ifx #1\expandafter \@firstoftwo
 \else \expandafter \@secondoftwo
 \fi
}%
\providecommand \natexlab [1]{#1}%
\providecommand \enquote  [1]{``#1''}%
\providecommand \bibnamefont  [1]{#1}%
\providecommand \bibfnamefont [1]{#1}%
\providecommand \citenamefont [1]{#1}%
\providecommand \href@noop [0]{\@secondoftwo}%
\providecommand \href [0]{\begingroup \@sanitize@url \@href}%
\providecommand \@href[1]{\@@startlink{#1}\@@href}%
\providecommand \@@href[1]{\endgroup#1\@@endlink}%
\providecommand \@sanitize@url [0]{\catcode `\\12\catcode `\$12\catcode
  `\&12\catcode `\#12\catcode `\^12\catcode `\_12\catcode `\%12\relax}%
\providecommand \@@startlink[1]{}%
\providecommand \@@endlink[0]{}%
\providecommand \url  [0]{\begingroup\@sanitize@url \@url }%
\providecommand \@url [1]{\endgroup\@href {#1}{\urlprefix }}%
\providecommand \urlprefix  [0]{URL }%
\providecommand \Eprint [0]{\href }%
\providecommand \doibase [0]{http://dx.doi.org/}%
\providecommand \selectlanguage [0]{\@gobble}%
\providecommand \bibinfo  [0]{\@secondoftwo}%
\providecommand \bibfield  [0]{\@secondoftwo}%
\providecommand \translation [1]{[#1]}%
\providecommand \BibitemOpen [0]{}%
\providecommand \bibitemStop [0]{}%
\providecommand \bibitemNoStop [0]{.\EOS\space}%
\providecommand \EOS [0]{\spacefactor3000\relax}%
\providecommand \BibitemShut  [1]{\csname bibitem#1\endcsname}%
\let\auto@bib@innerbib\@empty
\end{thebibliography}%


\begin{thebibliography}{99}
\bibitem{zwolak-2004}
M.~Zwolak and G.~Vidal, Mixed-State Dynamics in One-Dimensional 
Quantum Lattice Systems: A Time-Dependent Superoperator Renormalization Algorithm, 
\href{https://doi.org/10.1103/PhysRevLett.93.207205}{Phys. Rev. Lett. {\bf 93}, 207205 (2004)}. 

\bibitem{verstraete-2004}
F.~Verstraete, J.~J.~García-Ripoll, and J.~I.~Cirac, 
Matrix Product Density Operators: Simulation of Finite-Temperature and Dissipative Systems, 
\href{https://doi.org/10.1103/PhysRevLett.93.207204}{Phys. Rev. Lett. {\bf 93}, 207204 (2004)}.


\bibitem{hastings-2006}
M. B. Hastings, 
Solving gapped Hamiltonians locally, 
\href{https://doi.org/10.1103/PhysRevB.73.085115}{Phys. Rev. B {\bf 73}, 085115 (2006)}. 


\bibitem{prosen-2007}
T.~Prosen and M.~Znidaric, 
Is the efficiency of classical simulations of quantum dynamics related to integrability?, 
\href{https://doi.org/10.1103/PhysRevE.75.015202}{Phys.\ Rev.\ E {\bf 75}, 015202(R) (2007)}. 

\bibitem{znidaric-2008}
M.~Znidaric, T.~Prosen,  and I.~Pizorn, 
Complexity of thermal states in quantum spin chains, 
\href{https://doi.org/10.1103/PhysRevA.78.022103}{Phys.\ Rev.\ A {\bf 78}, 022103 (2008)}. 


\bibitem{molnar-2015}
A.~Molnar, N.~Schuch, F.~Verstraete, and J.~I.~Cirac, 
Approximating Gibbs states of local Hamiltonians efficiently with projected entangled pair states, 
\href{https://doi.org/10.1103/PhysRevB.91.045138}{Phys.\ Rev.\ B {\bf 91}, 045138 (2015)}. 


\bibitem{preskill-2018}
J.~Preskill, Quantum Computing in the NISQ era and beyond, 
\href{https://doi.org/10.22331/q-2018-08-06-79}{Quantum {\bf 2}, 79 (2018)}. 

\bibitem{zanardi}
P.~Zanardi, Entanglement of quantum evolutions, 
\href{http://dx.doi.org/10.1103/PhysRevA.63.040304}{Phys.\ Rev.\ A {\bf 63}, 040304(R) (2001)}. 

\bibitem{dubail-2017}
J.~Dubail, Entanglement scaling of operators: a conformal field theory approach, with a glimpse 
of simulability of long-time dynamics in $1+1d$,  
\href{https://doi.org/10.1088/1751-8121/aa6f38}{J.\ Physics A {\bf 50}, 234001 (2017)}. 

\bibitem{pizorn-2009}
I.~Pizorn and T.~Prosen, 
Operator space entanglement entropy in XY spin chains, 
\href{http://link.aps.org/doi/10.1103/PhysRevB.79.184416}{Phys.\ Rev.\ B {\bf 79}, 184416 (2009)}. 

\bibitem{hartmann-2009}
M.~J.~Hartmann, J.~Prior, S.~R.~Clark, and M.~B.~Plenio, 
Density Matrix Renormalization Group in the Heisenberg Picture, 
\href{http://link.aps.org/doi/10.1103/PhysRevLett.102.057202}{Phys. Rev. Lett. 102, 057202 (2009)}.

\bibitem{adm-2019}
V.~Alba, J.~Dubail, and M.~Medenjak
Operator Entanglement in Interacting Integrable 
Quantum Systems: the Case of the Rule 54 Chain, 
\href{https://doi.org/10.1103/PhysRevLett.122.250603}{Phys.\ Rev.\ Lett.\ 
	{\bf 122}, 250603 (2019)}. 

\bibitem{nahum-2017}
A.~Nahum, J.~Ruhman, S.~Vijay, and J.~Haah, 
Quantum entanglement growth under random unitary dynamics, 
\href{https://doi.org/10.1103/PhysRevX.7.031016}{Phys.\ Rev.\ X {\bf 7}, 031016 (2017)}.


\bibitem{nahum-2018}
A.~Nahum, S.~Vijay, and J.~Haah, 
Operator Spreading in Random Unitary Circuits, 
\href{https://doi.org/10.1103/PhysRevX.8.021014}{Phys.\ Rev.\ X {\bf 8}, 021014 (2018)}. 


\bibitem{keyser-2018}
C.~W.~von~Keyserlingk, T.~Rakovszky, F.~Pollmann, and S.~L.~Sondhi, 
Operator Hydrodynamics, OTOCs, and Entanglement Growth in Systems without Conservation Laws, 
\href{https://doi.org/10.1103/PhysRevX.8.021013}{Phys.\ Rev.\ X {\bf 8}, 021013 (2018)}. 


\bibitem{jonay-2018}
C.~Jonay, D.~Huse, and A.~Nahum, 
Coarse-grained dynamics of operator and state entanglement, 
\href{https://arxiv.org/abs/1803.00089}{arXiv:1803.00089}. 


\bibitem{khemani-2018}		
V.~Khemani, A.~Vishwanath, and D.~A.~Huse, 
Operator Spreading and the Emergence of Dissipative Hydrodynamics under Unitary Evolution with Conservation Laws, 
\href{https://doi.org/10.1103/PhysRevX.8.031057}{Phys. Rev. X 8, 031057 (2018)}. 

\bibitem{calabrese-2005}
	P.~Calabrese and J.~Cardy, Evolution of Entanglement Entropy in One-Dimensional Systems,
	 \href{http://dx.doi.org/10.1088/1742-5468/2005/04/P04010}{J. Stat. Mech. (2005) P04010}.


\bibitem{fagotti-2008}
M.~Fagotti and P.~Calabrese,
Evolution of entanglement entropy following a quantum quench: Analytic results for the XY chain in a transverse magnetic field, \href{https://doi.org/10.1103/PhysRevA.78.010306}{Phys. Rev. A 78, 010306 (2008)}.


\bibitem{alba-2016}
	 V.~Alba and P.~Calabrese, Entanglement and thermodynamics after a quantum quench in integrable systems,
	 \href{http://dx.doi.org/10.1073/pnas.1703516114}{PNAS {\bf 114}, 7947 (2017)}.

\bibitem{alba-2018}
	 V. Alba, P. Calabrese,
	 \href{https://dx.doi.org/110.21468/SciPostPhys.4.3.017}{SciPost Phys. 4, 017 (2018)}


\bibitem{bobenko-1993}
A.~Bobenko, M.~Bordemann, C.~Gunn, and  U.~Pinkall, 
On two integrable cellular automata, 
\href{https://doi.org/10.1007/BF02097234}{Commun.\ Math.\ Phys.\ {\bf 158}, 127 (1993)}. 


\bibitem{wigner-1955}
E.~P.~Wigner, Lower Limit for the Energy Derivative of the Scattering Phase Shift,  
\href{https://doi.org/10.1103/PhysRev.98.145}{Phys.\ Rev.\ {\bf 98}, 145 (1955)}


\bibitem{suppmat}
see Supplemental Material.

\bibitem{doyon-2018}
B.~Doyon, T.~Yoshimura, and J.-S.~Caux, 
Soliton Gases and Generalized Hydrodynamics, 
\href{https://doi.org/10.1103/PhysRevLett.120.045301}{Phys.\ Rev.\ Lett.\ {\bf 120}, 045301 (2018)}

\bibitem{klobas-2019}
K.~Klobas, M.~Medenjak, T.~Prosen, and M.~Vanicat, 
Time-Dependent Matrix Product Ansatz for Interacting Reversible Dynamics, 
\href{https://doi.org/10.1007/s00220-019-03494-5}{Commun.\ Math.\ Phys.\ {\bf 371}, 651 (2019)}. 

\bibitem{gopala-2018}
Operator growth and eigenstate entanglement in an interacting integrable Floquet system
S.~Gopalakrishnan, 
\href{https://doi.org/10.1103/PhysRevB.98.060302}{Phys. Rev. B {\bf 98}, 060302(R) (2018)}. 




\bibitem{sarang-2018}		
S.~Gopalakrishnan, D.~A.~Huse, V.~Khemani, and R.~Vasseur, 
Hydrodynamics of operator spreading and quasiparticle diffusion in interacting integrable systems, 
\href{https://doi.org/10.1103/PhysRevB.98.220303}{Phys.\ Rev.\ B {\bf 98}, 220303 (2018)}. 

\bibitem{jacopo}
J.~De~Nardis, D.~Bernard, and B.~Doyon, 
Hydrodynamic Diffusion in Integrable Systems, 
\href{https://doi.org/10.1103/PhysRevLett.121.160603}{Phys.\ Rev.\ Lett.\ {\bf 121} 160603 (2018)}. 

\bibitem{kiefer-2020}
M.~Kiefer-Emmanouilidis, R.~Unanyan, J.~Sirker, and M.~Fleischhauer, 
Bounds on the entanglement entropy by the number entropy in non-interacting fermionic systems, 
\href{https://arxiv.org/abs/2003.03112}{arXiv:2003.03112}. 


\bibitem{spohn-2018}
H.~Spohn, Interacting and noninteracting integrable systems, 
\href{https://doi.org/10.1063/1.5018624}{J.\ Math.\ Phys.\ {\bf 59}, 091402 (2018)}.

\bibitem{uli}
U.~Schollw\"ock, 
The density-matrix renormalization group in the age of matrix product states, 
\href{https://doi.org/10.1016/j.aop.2010.09.012}{Annals of Physics {\bf 326}, 96 (2011)}. 

\bibitem{paeckel-2019}
S.~Paeckel, T.~K\"ohler, A.~Swoboda, S.~R.~Manmana, U.~Schollw\"ock, and C.~Hubig, 
Time-evolution methods for matrix-product states, 
\href{https://doi.org/10.1016/j.aop.2019.167998}{Annals of Physics {\bf 411}, 167998 (2019)}. 

\bibitem{itensor}
Simulations are performed by using the ITensor library, 
\href{http://itensor.org}{itensor.org}. 

\bibitem{ilievski-2018}
E.~Ilievski, J.~De~Nardis, M.~Medenjak, and T.~Prosen, 
Superdiffusion in One-Dimensional Quantum Lattice Models, 
\href{https://doi.org/10.1103/PhysRevLett.121.230602}{Phys.\ Rev.\ Lett.\ {\bf 121}, 230602 (2018)}.


\bibitem{gopala-2019}
S.~Gopalakrishnan and R.~Vasseur, 
Kinetic Theory of Spin Diffusion and Superdiffusion in 
$XXZ$ Spin Chains, 
\href{https://doi.org/10.1103/PhysRevLett.122.127202}{Phys. Rev. Lett. {\bf 122}, 127202 (2019)}. 

\bibitem{ljubotina-2017}
M.~Ljubotina, M.~Žnidarič, and T.~Prosen, 
Spin diffusion from an inhomogeneous quench in an integrable system, 
\href{}{Nature Communications {\bf 8}, 16117 (2017)}. 


\bibitem{muth-2011}
D.~Muth, R.~G.~Unanyan, and M.~Fleischhauer, 
Dynamical Simulation of Integrable and Nonintegrable Models in the Heisenberg Picture, 
\href{https://doi.org/10.1103/PhysRevLett.106.077202}{Phys.\ Rev.\ Lett.\ {\bf 106}, 077202 (2011)}. 

\bibitem{bertini-2019}
B.~Bertini, P.~Kos, and T.~Prosen, 
Entanglement Spreading in a Minimal Model of Maximal Many-Body Quantum Chaos, 
\href{https://doi.org/10.1103/PhysRevX.9.021033}{Phys. Rev. X {\bf 9}, 021033 (2019)}.

\bibitem{bertini-2020}
B.~Bertini, P.~Kos, and T.~Prosen, 
Operator Entanglement in Local Quantum Circuits I: Chaotic Dual-Unitary Circuits, 
\href{https://doi.org10.21468/SciPostPhys.8.4.067}{SciPost Phys. {\bf 8}, 067 (2020)}. 

\bibitem{bertini-2020a}
B.~Bertini, P.~Kos, and T.~Prosen, 
Operator Entanglement in Local Quantum Circuits II: Solitons in Chains of Qubits, 
\href{https://doi.org/10.21468/SciPostPhys.8.4.068}{SciPost Phys. {\bf 8}, 068 (2020)}. 


\bibitem{klobas-2020}
K.~Klobas and T.~Prosen, 
Space-like dynamics in a reversible cellular automaton, 
\href{https://arxiv.org/abs/2004.01671}{arXiv:2004.01671}. 


\bibitem{klobas-2020a}
K.~Klobas, M.~Vanicat, J.~P.Garrahan, and T.~Prosen, 
Matrix product state of multi-time correlations,
\href{https://doi.org/10.1088/1751-8121/ab8c62}{J.\ Phys.\ A {\bf 10}, 1088 (2020)}. 

\bibitem{dupont-2020}
M.~Dupont and J.~E.~Moore, 
Universal spin dynamics in infinite-temperature one-dimensional quantum magnets, 
\href{https://doi.org/10.1103/PhysRevB.101.121106}{Phys. Rev. B {\bf 101}, 121106(R)}.


\bibitem{jacopo-1}
J.~De~Nardis, M.~Medenjak, C.~Karrasch, and E.~Ilievski, 
Universality Classes of Spin Transport in One-Dimensional Isotropic Magnets: The Onset of Logarithmic Anomalies, 
\href{https://doi.org/10.1103/PhysRevLett.124.210605}{Phys.\ Rev.\ Lett.\ {\bf 124}, 210605 (2020)}. 




\bibitem{oganesyan-2007}
V.~Oganesyan and D.~Huse, Localization of interacting fermions at high temperature, 
\href{https://doi.org/10.1103/PhysRevB.75.155111}{Phys.\ Rev.\ B {\bf 75}, 155111 (2007)}. 


\bibitem{atas-2013}
Y.~Y.~Atas, E.~Bogomolny, O.~Giraud, and G.~Roux, 
The distribution of the ratio of consecutive level spacings in random matrix ensembles, 
\href{https://doi.org/10.1103/PhysRevLett.110.084101}{Phys.\ Rev.\ Lett.\ {\bf 110}, 084101, (2013)}. 

%

%

%

%

%

\end{thebibliography}
\end{document}